\documentstyle[12pt]{article}
\begin{document}
\textheight=22 true cm
\textwidth=15 true cm
\normalbaselineskip=24 true pt
\normalbaselines
\topmargin -0.5 true in
\bibliographystyle{unsrt}
\def\sl{\em}
\def\vslash{v\!\!\!/}
\def\vpslash{{v^\prime}\!\!\!\!/}
\def\lslash{l\!\!\!/}
\def\epslash{\epsilon\!\!\!/}
\def\be{\begin{equation}}
\def\ee{\end{equation}}
\def\bea{\begin{eqnarray}}
\def\eea{\end{eqnarray}}

\setcounter{page}{0}
\thispagestyle{empty}

\begin{flushright}
{\sf SINP-TNP/96-19}
\end{flushright}

\begin{center}
{\Large \bf Non-relativistic Model for the \\
   Semileptonic ${\unboldmath \Lambda_b \to \Lambda_c}$ Decay}\\[5mm]
{\large\sf Debrupa Chakraverty\footnote{Electronic address:
 rupa@tnp.saha.ernet.in}, Triptesh De\footnote{Electronic address:
 td@tnp.saha.ernet.in}, Binayak Dutta-Roy\footnote{Electronic
 address: bnyk@tnp.saha.ernet.in}}\\
and\\[5mm]
{\large\sf K. S. Gupta\footnote{Electronic address:
gupta@tnp.saha.ernet.in}}\\
{\sl Theory Group, Saha Institute of Nuclear Physics,\\
1/AF Bidhannagar, Calcutta - 700 064, India}\\[5mm]
\end{center}

\begin{abstract}
 We calculate the decay width for $\Lambda_b \to \Lambda_c e {\bar \nu}$
 in the frame work of a nonrelativistic quark (NRQ) model of heavy baryons
where the light quarks play the role of spectators. Our calculation does not
make an explicit use of the heavy quark symmetry. The branching ratio for
the above process as calculated here agrees reasonably well with the
experimental value. 
\end{abstract}
\bigskip

\newpage
\centerline {\bf I. INTRODUCTION}
\bigskip

Non-relativistic quark (NRQ) models have played an important role in
understanding various properties of hadrons. In the recent past they
have been used in the literature to
study semileptonic decay of heavy mesons \cite{isgw}. Such calculations
can partly be justified from the fact that the heavy quark in a heavy hadron
acts as a source of static colour field. 

In this Letter we extend
such calculations to the case of baryons containing a single heavy quark.
Specifically we look at the semileptonic decay of 
  $\Lambda_b \to \Lambda_c e {\bar \nu}$. 
The study of this decay provides an opportunity to measure the CKM matrix
element $V_{cb}$ and also to test the limits of the NRQ model calculations
Even though we are dealing with systems containing heavy quarks, no explicit
use of the heavy quark symmetry is used in our calculation.
This enables 
us to express quantities of physical interest 
explicitly in terms of the masses of
heavy quarks (and other parameters of the model.) Our calculation thus by
construction incorporates some of the finite mass corrections.

 This paper is organised as follows: the next section sets forth the review 
 of the underlying kinematics. In section 
 III, we discuss the NRQ model as a tool for calculating the form factors.
Section IV  concludes the paper with the results and discussions.

\bigskip 

\centerline{\bf II. KINEMATICS }
\bigskip
 The matrix element for the process
  $\Lambda_b \to \Lambda_c e {\bar \nu}$ is given by,
\be
A_{s^{\prime} s} = {G_F\over \sqrt{2}} V_{cb} 
L^{\mu} {H_{\mu}}^{(s^{\prime}s)}, 
\ee
 with the 
 leptonic and hadronic parts being,
 \be
 L^{\mu} = {\bar u}_e (p)\gamma^{\mu}(1-\gamma_5)v_{\nu}(p^{\prime}),
\ee
\be
{H^{(s^{\prime} s)}}^{\mu} = 
\langle K,s^{\prime} \vert {J_{had}}^{\mu} \vert P,s \rangle. 
\end{equation}
$P, K, p$ and $p^{\prime}$ are the four momenta corresponding to
$\Lambda_b, \Lambda_c, e$ and $\bar \nu$ respectively.
The corresponding decay rate
 \cite{single, pdg} is given by
 \be
{{d \Gamma_{s^{\prime} s}} \over {dy d\Omega_e d{\tilde \Omega_{\Lambda_c}}}}
 = {1\over 2} {
\vert {\tilde K} \vert 
\over {(4 \pi)^5}} \vert A_{s^{\prime} s} \vert^2,
\ee
where $y= {q^2\over {m_{\Lambda_b}}^2}$ is a dimensionless kinematic
variable. The four momentum of the virtual $W$ is given by
$q= p + p^{\prime}= P - K$.
 With both the parent and daughter baryons possessing 
spin ${1\over 2}$ 
the  spin components s, $s^{\prime}$ of $\Lambda_b$ and $\Lambda_c$
can  be either $+{1\over 2}$ or $-{1\over 2}$. 
 The  solid angle of the electron in the $e {\bar \nu} $ centre of 
mass frame ( $ e {\bar \nu}$ frame) is denoted by 
 $d{ \Omega}_e$  
while $d{\tilde \Omega}_{\Lambda_c}$ denotes 
the solid angle of the daughter 
baryon $\Lambda_c$ in the rest frame of parent baryon $\Lambda_b$.
 In the parent rest frame the quantities are denoted by a tilde and  in 
the  $e {\bar \nu} $ frame  without the tilde. 
 
In  the $e {\bar \nu} $  frame, the energy and three momenta  
$ (E_{\Lambda_b}, \vec P)$ and 
$(E_{\Lambda_c}, \vec K)$  of $\Lambda_b$ 
and $\Lambda_c$ respectively, are easily determined to be 
\be
E_{\Lambda_b} = {m_{\Lambda_b}\over {2 \sqrt{y}}} 
(1 - {{m_{\Lambda_c}}^2\over {m_{\Lambda_b}}^2} +y),
\ee
\be
E_{\Lambda_c} = {m_{\Lambda_b}\over {2 \sqrt{y}}} 
(1 - {{m_{\Lambda_c}}^2\over
 {m_{\Lambda_b}}^2} -y),
\ee
$$\vert P \vert =\vert K \vert = {k\over {\sqrt{y}}},$$
 while in parent rest frame $(E_{\Lambda_b} = m_{\Lambda_b}, \vec P = 0)$,
 the  energy and three momenta of daughter baryon $\Lambda_c$ are
\be
{\tilde E_{\Lambda_c}} = {m_{\Lambda_b}\over 2} 
(1 + {{m_{\Lambda_c}}^2\over {m_{\Lambda_b}}^2} -y),
\ee
\be
\vert {\tilde K} \vert \equiv k
 = {m_{\Lambda_b}\over 2} [(1 - {{m_{\Lambda_c}}^2\over
 {m_{\Lambda_b}}^2} -y)^2 - {{4 {m_{\Lambda_c}}^2}\over 
{m_{\Lambda_b}}^2} y]^{1\over 2}.
\ee

 In terms of form factors, the hadronic matrix elements of the current 
${J_{had}}^{\mu}= V^{\mu} - A^{\mu}$ are written as,
 \be
 \langle K,s^{\prime} \vert V^{\mu} \vert 
 P,s \rangle = {{\bar u}_{\Lambda_c}}^{s^{\prime}}[ g(q^2)\gamma^{\mu}
+ g_+(q^2)(P+K)^{\mu}+g_-(q^2)(P-K)^{\mu}]u^s_{\Lambda_b},
\ee
 \be
 \langle K,s^{\prime} \vert A^{\mu} \vert 
 P,s \rangle = {{\bar u}_{\Lambda_c}}^{s^{\prime}}[ a(q^2)\gamma^{\mu}
 \gamma_5 +a_+(q^2)(P+K)^{\mu}\gamma_5+a_-(q^2)(P-K)^{\mu}\gamma_5]
u^s_{\Lambda_b}.
\ee
 Here 
 $u_{\Lambda_b}(u_{\Lambda_c})$
 is the spinor associated with $\Lambda_b(\Lambda_c)$. 

 The decay width for the process under consideration is 
given by the expression  \cite{single}
 \be
 \Gamma = \int^{y_{max}}_{y_{min}} {{G_F^2 \vert V_{cb} \vert^2 k
 { m_{\Lambda_b}}^2 y}\over {96 \pi^3}}
(\vert H_+ \vert^2 + \vert H_- \vert^2
 + \vert H_0 \vert^2),
\ee
 where
\be
H_{\pm} = \pm (a F_0 \mp g F_-),
\ee
\be
H_0 = \{ [ 2 a (1 - {1\over 2}F_0) - 2 {k\over \sqrt{y}} a_+ F_-]^2+
 (2{k\over \sqrt{y}} g_+ F_0 +g F_+)^2 \}^{1\over 2},
\ee
with
\be
F_{\pm}= [{{(E_{\Lambda_b}+m_{\Lambda_b})( E_{\Lambda_c}+m_{\Lambda_c})}
\over {4 m_{\Lambda_b} m_{\Lambda_c}}}]^{1\over 2}[{k \over {\sqrt{y}
 (E_{\Lambda_c}+m_{\Lambda_c})}} \pm {k \over {\sqrt{y} (E_{\Lambda_b}
+m_{\Lambda_b})}}],
\ee
and
\be
F_0 = [{{(E_{\Lambda_b}+m_{\Lambda_b})( E_{\Lambda_c}+m_{\Lambda_c})} 
\over {4 m_{\Lambda_b} m_{\Lambda_c}}}]^{1\over 2}[ 1 - {k^2\over {y
(E_{\Lambda_b}+m_{\Lambda_b})( E_{\Lambda_c}+m_{\Lambda_c})}}].
\ee
Within the kinematically allowed region, the lower ($ y_{min}$) and
upper ($y_{max}$) limits of y are 0 and $(1-{m_{\Lambda_c}\over 
m_{\Lambda_b}})^2$ respectively (neglecting the electron mass).

	The form factors given above can be evaluated within the framework
of some model. We employ the NRQ model for this purpose which is described
next.

\eject
\newpage
\bigskip
\centerline{\bf III. NONRELATIVISTIC QUARK MODEL AND FORMFACTORS}
\bigskip

In the NRQ model, the 
parent baryon $\Lambda_b$ 
contains the  heavy quark b and two light quarks u and d
 (having nearly equal masses).
The light quarks are taken to behave as spectators. Similarly
 the daughter baryon 
consists of the  charmed quark and the same two spectator light quarks.
 The spatial coordinates 
of the three quarks in $\Lambda_b$ are  denoted by $\vec 
 r_b$, $\vec r_1$ and $\vec r_2$, while those in $\Lambda_c$ are denoted 
 by $\vec r_c$, $\vec {r_1^{\prime}}$,
  $\vec {r_2^{\prime}}$. It is convenient to introduce 
 the Jacobi coordinates $\vec R$, $\vec \rho$
 and $\vec \lambda$ defined as \cite{bhaduri}
 \be
\vec R ={ {m_1 \vec r_1 + m_2 \vec r_2 + m_b \vec r_b}\over
 {m_1+m_2+m_b}},
\ee
\be
\vec \rho = {1\over \sqrt{2}}(\vec r_1 - \vec r_2),
\ee
\be
\vec \lambda =  {1\over \sqrt{6}}(\vec r_1 + \vec r_2 - 2 \vec r_b).
\ee
We take $ m_1 = m_2 = m_u$ ( 
mass of each light quark contained in the spectator pair.)
The corresponding canonically conjugate momenta
  $\vec P$, $\vec p_{\rho}$
 and  $\vec p_{\lambda}$ are respectively,
\be
\vec P = \vec p_1 + \vec p_2 + \vec p_b,
\ee
\be
\vec p_{\rho} = {1\over {\sqrt{2}}}(\vec p_1 - \vec p_2),
\ee
\be
 \vec p_{\lambda} = \sqrt{3\over 2}[{m_b\over {\bar m_{\Lambda_b}}}
(\vec p_1 +\vec p_2) - {{2m_u}\over 
 {\bar m_{\Lambda_b}}}\vec p_b],
\ee
 where $\vec p_1$ and $\vec p_2$ are momenta of light quarks
 and $\bar m_{\Lambda_b} = 2m_u +m_b$.  
We have taken $m_{\Lambda_b}$ (the physical
 mass of $\Lambda_b$) =  $\bar m_{\Lambda_b}$.

In this model, the normalized $\Lambda_b$ baryon state vector
 is written as,
 \bea
\vert \Lambda_b (P,s) \rangle & = & \sqrt{2 m_{\Lambda_b}} \sum 
C^s_{s_1,s_2,s_b}
\int d^3 \vec p_{\lambda}
 d^3 \vec p_{\rho} \phi_{\Lambda_b}(\vec p_{\lambda}, \vec p_{\rho})
\nonumber\\
& {} & 
 \vert u(p_1, s_1); d(p_2,s_2); b(p_b,s_b)\rangle,
\eea
 where $\phi_{\Lambda_b}(\vec p_{\lambda}, \vec p_{\rho})$ is the momentum 
space wave function, satisfying the normalisation condition,
\be
\int d^3 \vec p _{\lambda} d^3 \vec p_{\rho} \vert \phi_{\Lambda_b}
(\vec p_{\lambda}, \vec p_{\rho}) \vert^2 =1.
\ee
${C^s}_{s_1,s_2.s_b}$ is the Clebsch-Gordon coefficient for  combining  
three spin-$1\over 2$ 
constituent quarks into a spin ${1\over 2}$ baryon with the 
spin component s along the z axis.

 The flavor and spin part of $\Lambda_b$ is given by
 \be
\vert \Lambda_b, s\rangle =\vert u d [I=0, S=0]; b\rangle {\cal S}_s(1),
\ee
 where
\bea
{ \cal S}_s(1) & = &  {1\over \sqrt{2}} (\vert +-+\rangle - \vert -++\rangle)
 {\quad} {\rm with} {\quad} s= +{1\over 2},\nonumber\\
 & = & {1\over \sqrt{2}} (\vert +--\rangle - \vert -+-\rangle)
 {\quad} {\rm with} {\quad} s= -{1\over 2}\nonumber
\eea
following  the ordering $\vert s_1 s_2 s_b\rangle$;

Similarly,
 \bea
\vert \Lambda_c (K,s^{prime}) \rangle & = & \sqrt{2 m_{\Lambda_c}} \sum 
C^{s^{\prime}}_{s_1,s_2,s_c}
\int d^3 \vec p_{\lambda}^{\prime}
 d^3 \vec p_{\rho}^{\prime} \phi_{\Lambda_c}(\vec p_{\lambda}^{\prime}, 
\vec p_{\rho}^{\prime})
\nonumber\\
& {} & 
 \vert u(p_1, s_1); d(p_2,s_2); c(p_c,s_c)\rangle,
\eea

 The Hamiltonian that we consider is given by,
 \be
H= \sum_i {{p_i}^2\over {2m_i}} +\sum_{i<j}({1\over 2} a_1 r_{ij}
 + {a_2} -{{2\alpha_s}\over {3r_{ij}}}).
\ee
 Here the ``${1\over 2}$ rule" 
i.e. $V_{qq} = {1\over 2} V_{q\bar q}$ \cite{richard} 
  is applied. 
The parameter $a_1 = 0.18 GeV^2$ \cite{isgw} 
is fixed from the meson mass spectrum. 
  It is obvious that the wavefunctions should be independent of 
the parameter $a_2$.

The  Hamiltonian (Eqn. 25) is best re-expressed  \cite{bhaduri} 
in terms of Jacobi coordinates introduced in Eqns 16-18.  Thus
\bea
H & = & {p^2_{\rho} \over {2 m_{\rho}}}+
{p^2_{\lambda} \over {2 m_{\lambda}}}+
{ P^2\over {2 m}} +{1\over 2}a_1\vert \sqrt{2} {\vec \rho}\vert \nonumber\\
 & {} & +  {1\over 2} a_1\vert {1\over \sqrt{2}}(\vec \rho +\sqrt{3} 
\vec \lambda)\vert
 + {1\over 2} a_1\vert {1\over \sqrt{2}}
(\vec \rho - \sqrt{3} \vec \lambda )\vert
 +a_2 \nonumber\\
& {} &-{2\over 3}\alpha_s\{ {1\over {\vert \sqrt{2} \vec \rho\vert}} 
+{\sqrt{2}
 \over {\vert (\vec \rho +\sqrt{3} \vec \lambda)\vert}} +
+{\sqrt{2}
 \over {\vert (\vec \rho -\sqrt{3} \vec \lambda)\vert}}\}, 
 \eea
 where $m= m_{\Lambda_b}$, $m_{\rho} = m_1 = m_2=m_u$ and
 $ m_{\lambda} = {{3 m_u m_b}\over {2 m_u +m_b}}$. 
  The values of the mass parameters we shall use are 
$m_u = 0.35GeV$, $m_c= 1.80 GeV$, $m_b =
 5.00 GeV$ and the coupling strength relevant to the b \& c hadrons 
 should be taken to be $\alpha_s = 0.5$. 

The wave function $\phi_{\Lambda_b}( \vec {p_{\rho}}, 
 \vec {p_{\lambda}})$ is an eigenfunction of the above H given by Eqn.26.
Employing a 
  variational approach with 
 trial wavefunctions taken as the product of normalised Gaussians in 
the variables $\rho$ and $\lambda$,
\be
\phi(\rho, \lambda) = ({\alpha_1\over \pi})^{3\over 4}
 ({\alpha_2\over \pi})^{3\over 4} exp(-\alpha_1 \rho^2/2)
  exp(-\alpha_2 \lambda^2/2).
\ee
and with the $\Lambda_c$ baryon wavefunction  expressed with analogous
 variables written with primes,
 the value of the parameters $\alpha_1$ and
 $\alpha_2$  minimizing the ground state energies for $\Lambda_b$ and
 $\Lambda_c$ are found to be
 $$\alpha_1^b = 0.264{\quad} {\rm and}{\quad}\alpha_2^b = 0.653{\quad}
 {\rm for}{\quad}\Lambda_b,$$
 $$\alpha_1^c = 0.261{\quad} {\rm and}{\quad}\alpha_2^c = 0.568{\quad}
 {\rm for}{\quad}\Lambda_c.$$

In the rest frame of the $\Lambda_b$ baryon, ($\vec P = 0$)
\be
\vec p_c = \vec p_b + \vec {\tilde K} =  
 \vec {\tilde K} -\sqrt{2\over 3} \vec p_{\lambda}
\ee
 with spectator approximation ($\vec {p_1}^{\prime} = \vec p_1$ 
and $\vec {p_2}^{\prime} = \vec p_2$),
\be
 \vec {p_{\rho}}^{\prime} =\vec  p_{\rho},
\ee
\be
 \vec {p_{\lambda}}^{\prime} = \vec p_{\lambda } - \sqrt{3\over 2} {2m_u
 \over m_{\Lambda_c}} \vec {\tilde K}.
\ee
 Using these baryon state vectors, the hadronic matrix
elements are found to be
 \bea
\langle \vec {\tilde K}, s^{\prime} \vert J^{\mu} \vert \vec 0, s 
\rangle & = &
\sum {C^{s^{\prime}}}_{s_1, s_2, s_c}
 {C^s}_{s_1,s_2,s_b}
 \int d^3 \vec {{p_{\rho}}^{\prime}} d^3 \vec {{p_{\lambda}}^{\prime}}
d^3 \vec p_{\rho} d^3 \vec p_{\lambda} \nonumber\\
 & { } & \delta^{(3)} (
 \vec {p_{\lambda}}^{\prime} - \vec p_{\lambda } + \sqrt{3\over 2} 
{2m_u \over  m_{\Lambda_c}} \vec {\tilde K})
 \delta^{(3)}( \vec {p_{\rho}}^{\prime} -\vec  p_{\rho})\nonumber\\
& { }  & {\phi_{\Lambda_c}}^*(
 \vec {p_{\rho}}^{\prime}, 
 \vec {p_{\lambda}}^{\prime}) 
 {\phi_{\Lambda_b}}(
 \vec {p_{\rho}}, 
 \vec {p_{\lambda}}) \nonumber\\
& { }  & \bar c (\vec p_c, s_c) \gamma^{\mu} (1-\gamma_5) b(\vec 
 p_b, s_b).
\eea
 Writing the relevant matrix elements for the temporal and spatial 
components of the currents separately we get 
\be
{{\tilde V}_{s^{\prime}, s}}^0 =
(4 m_{\Lambda_b} m_{\Lambda_c})^{1\over 2} \int 
d^3 \vec p_{\rho} d^3 \vec p_{\lambda} 
 {\phi_{\Lambda_c}}^*(
 \vec {p_{\rho}}, 
 \vec {p_{\lambda}}  
 - \sqrt{3\over 2} {2m_u
 \over  m_{\Lambda_c}} \vec {\tilde K})
 {\phi_{\Lambda_b}}(
 \vec {p_{\rho}}, 
 \vec {p_{\lambda}}) 
 {\chi_{s^{\prime}}}^{\dagger}
\varphi_s,
\ee 
\bea
 {{\vec {\tilde V}}_{s^{\prime}, s}} & = & 
(4 m_{\Lambda_b} m_{\Lambda_c})^{1\over 2} \int 
d^3 \vec p_{\rho} d^3 \vec p_{\lambda} \nonumber\\
 & { } & {\phi_{\Lambda_c}}^*(
 \vec {p_{\rho}}, 
 \vec {p_{\lambda}}  
 - \sqrt{3\over 2} {2m_u
 \over  m_{\Lambda_c}} \vec {\tilde K})
 {\phi_{\Lambda_b}}(
 \vec {p_{\rho}}, 
 \vec {p_{\lambda}})\nonumber\\
& { } &  
 {\chi_{s^{\prime}}}^{\dagger} \{ 
 i {{\hat {\sigma} \times
 ({\vec {\tilde K}}- \sqrt{2\over 3}\vec p_{\lambda})}
\over {2 m_c}}
+
{{({\vec {\tilde K}}- \sqrt{2\over 3}\vec p_{\lambda})}
\over {2 m_c}}\nonumber\\
& { } &  +i {{\hat {\sigma} \times
  \sqrt{2\over 3}\vec p_{\lambda}}
\over {2 m_b}} - 
  {{\sqrt{2\over 3}\vec p_{\lambda}}
\over {2 m_b}} \} 
 \varphi_s,
\eea 
\bea
{{\tilde A}_{s^{\prime}, s}}^0  & = & 
(4 m_{\Lambda_b} m_{\Lambda_c})^{1\over 2} \int 
d^3 \vec p_{\rho} d^3 \vec p_{\lambda} \nonumber\\
& { } &  {\phi_{\Lambda_c}}^*(
 \vec {p_{\rho}}, 
 \vec {p_{\lambda}}  
 - \sqrt{3\over 2} {2m_u
 \over  m_{\Lambda_c}} \vec {\tilde K})
 {\phi_{\Lambda_b}}(
 \vec {p_{\rho}}, 
 \vec {p_{\lambda}}) \nonumber\\
& { } & 
 {\chi_{s^{\prime}}}^{\dagger}
\{ 
  {{\hat {\sigma} \cdot
 ({\vec {\tilde K}}- \sqrt{2\over 3}\vec p_{\lambda})}
\over {2 m_c}}
 - {{\hat {\sigma} \cdot
  \sqrt{2\over 3}\vec p_{\lambda}}
\over {2 m_b}} \} 
\varphi_s,
\eea 
\bea
 {{\vec {\tilde A}}_{s^{\prime}, s}} & = & 
(4 m_{\Lambda_b} m_{\Lambda_c})^{1\over 2} \int 
d^3 \vec p_{\rho} d^3 \vec p_{\lambda} \nonumber\\
& { } & {\phi_{\Lambda_c}}^*(
 \vec {p_{\rho}}, 
 \vec {p_{\lambda}}  
 - \sqrt{3\over 2} {2m_u
 \over m_{\Lambda_c}} \vec {\tilde K})
 {\phi_{\Lambda_b}}(
 \vec {p_{\rho}}, 
 \vec {p_{\lambda}})\nonumber\\
& { } & 
(1 + {{\tilde K}^2\over {8 {m_c}^2}}) 
{\chi_{s^{\prime}}}^{\dagger} {\hat {\sigma}}
\varphi_s.
\eea 

In order to extract the form factors appearing in Eqns. 9 - 10, we write
their temporal and spatial components separately   
 in the parent rest
 frame with $\vert {\tilde K} \vert << m_{\Lambda_c}$ 
( non-relavistic limit)
for comparison with results from NRQ model :
 \be
{{\tilde V}_{s^{\prime}, s}}^0 = {\chi_{s^{\prime}}}^{\dagger}[g(q^2)
 + (m_{\Lambda_b} +m_{\Lambda_c}) g_+(q^2) +
  (m_{\Lambda_b} -m_{\Lambda_c}) g_-(q^2) ]\varphi_s
\ee
 \be
 {{\vec {\tilde V}}_{s^{\prime}, s}} = 
{\chi_{s^{\prime}}}^{\dagger}[\{{g(q^2)
\over {2 m_{\Lambda_c}}} + g_+(q^2)-
 g_-(q^2)\}{\vec {\tilde K}} +i {{\hat {\sigma} \times {\vec {\tilde K}}}
\over {2 m_{\Lambda_c}}}
 g(q^2)]\varphi_s
\ee
\be
{{\tilde A}_{s^{\prime}, s}}^0 = {\chi_{s^{\prime}}}^{\dagger}[a(q^2)
 - (m_{\Lambda_b} +m_{\Lambda_c}) a_+(q^2) -
  (m_{\Lambda_b} -m_{\Lambda_c}) a_-(q^2) ]
  \{{{\hat {\sigma} \cdot {\vec {\tilde K}}}\over {2 m_{\Lambda_c}}}\}
\varphi_s
\ee
\be
 {{\vec {\tilde A}}_{s^{\prime}, s}} = 
{\chi_{s^{\prime}}}^{\dagger}[\{a(q^2)
{\hat {\sigma}}
[1 + {{\tilde K}^2\over {8 {m_{\Lambda_c}}^2}}]\} -
( a_+(q^2) -a_-(q^2)) \{{{\hat {\sigma} \cdot {\vec {\tilde K}}}\over
 {2 m_{\Lambda_c}}}\}{\tilde K}]
\varphi_s
\ee
Here $\varphi_s$ and $\chi_{s^{\prime}}$ are two component Pauli spinors
 along z (parent spin direction) and $\vec \epsilon$ (daughter 
 polarization direction), the latter being taken to be arbitrary.
 It is to be noted that in the 
$e {\bar \nu}$ rest frame the parent and the daughter 
 have equal and opposite three momenta.

Inserting the Fourier transform 
of the above wavefunctions $[\phi(\vec p_{\rho},
 \vec p_{\lambda})]$ in the expressions for the matrix elements of the 
 vector and axial vectors 
currents [Eqns. 32 - 35] and then comparing with the
 matrix elements expressed 
in terms of form factors [Eqns. 36 - 39], the 
following relations emerge :
\be
g+(m_{\Lambda_b} +m_{\Lambda_c}) g_+ +
(m_{\Lambda_b} -m_{\Lambda_c}) g_- =
I (4 m_{\Lambda_c} m_{\Lambda_b})^{1\over 2}
 exp(-{{3 m_u^2 {\tilde K}^2} \over { (\alpha_2^c + \alpha_2^b) 
m^2_{\Lambda_c}}}),
\ee
 \bea
{g\over {2 m_{\Lambda_c}}} +g_+ - g_-  & = & 
 {I \over { m_c}}( m_{\Lambda_c} m_{\Lambda_b})^{1\over 2}
 exp(-{{3 m_u^2 {\tilde K}^2} \over { (\alpha_2^c + \alpha_2^b) 
m^2_{\Lambda_c}}})\nonumber\\
& { } & [{1 - 
{{ 2 m_u \alpha_2^b}\over {m_{\Lambda_c}m_b(\alpha_2^c 
+\alpha_2^b)}}(m_b +m_c)}],
\eea
 \bea
{g\over {2 m_{\Lambda_c}}}   & = & 
 {I\over { m_c}} ( m_{\Lambda_c} m_{\Lambda_b})^{1\over 2}
 exp(-{{3  m_u^2 {\tilde K}^2} \over { (\alpha_2^c + \alpha_2^b) 
m^2_{\Lambda_c}}})\nonumber\\
& { } & [{1 - 
{{ 2 m_u \alpha_2^b}\over {m_{\Lambda_c}m_b(\alpha_2^c 
+\alpha_2^b)}}(m_b -m_c)}],
\eea
\bea
 & { } & a-(m_{\Lambda_b} +m_{\Lambda_c}) a_+ -
(m_{\Lambda_b} -m_{\Lambda_c}) a_-   =  
  m_{\Lambda_c} {I \over m_c}
  (4 m_{\Lambda_c} m_{\Lambda_b})^{1\over 2}\nonumber\\
 & { } &  exp(-{{3 m_u^2 {\tilde K}^2} \over { (\alpha_2^c + \alpha_2^b) 
m^2_{\Lambda_c}}})
 [{1 - 
{{ 2 m_u \alpha_2^b}\over {m_{\Lambda_c}m_b(\alpha_2^c 
+\alpha_2^b)}}(m_b +m_c)}],
\eea
\be
a =
I (4 m_{\Lambda_c} m_{\Lambda_b})^{1\over 2}
 exp(-{{3 m_u^2 {\tilde K}^2} \over { (\alpha_2^c + \alpha_2^b) 
m^2_{\Lambda_c}}}),
\ee
\be
a_+ - a_- = 0,
\ee
where the overlap integral $I$  is given by
 \bea
I & = & 
 ({1 \over {\alpha_1^b \pi}})^{3\over 4}
 ({1 \over {\alpha_2^b \pi}})^{3\over 4}
 ({1 \over {\alpha_1^c \pi}})^{3\over 4}
 ({1 \over {\alpha_2^c \pi}})^{3\over 4}
  \int d^3 {\vec p}_{\rho} d^3 {\vec {\bar p}}_{\lambda} \nonumber\\
 & { } & exp[- ( {{\alpha_1^b + \alpha_1^c}
\over {2 \alpha_1^c \alpha_1^b}})
 p_{\rho}^2]
  exp[- ( {{\alpha_2^b + \alpha_2^c}\over {2 \alpha_2^c \alpha_2^b}})
 {\bar p}_{\lambda}^2]\nonumber\\
 & = & {{8 {\alpha_1^b}^{3\over 4} {\alpha_2^b}^{3\over 4}
  {\alpha_1^c}^{3\over 4} {\alpha_2^c}^{3\over 4}}\over 
 {{(\alpha_1^b +\alpha_1^c)}^{3\over 2}
 {(\alpha_2^b +\alpha_2^c)}^{3\over 2}}},
 \eea
with 
$${\vec {\bar p}}_{\lambda} = \vec p_{\lambda} - \sqrt{3\over 2} 
{{2m_u \alpha_2^b}\over {m_{\Lambda_c}(\alpha_2^c +\alpha_2^b)}} 
{\vec {\tilde K}}.$$
 From [eqs. 40 - 45] the form factors are obtained as functions of 
y;
\bea
g    & = & 
  m_{\Lambda_c} {I\over m_c} (4 m_{\Lambda_c} m_{\Lambda_b})^{1\over 2}
 exp[-\alpha \{( \beta -y)^2 - \gamma y\}] 
\nonumber\\
& { } & [{1 - 
{{ 2 m_u \alpha_2^b}\over {m_{\Lambda_c}m_b(\alpha_2^c 
+\alpha_2^b)}}(m_b -m_c)}],
\eea
\bea
g_{\pm}    & = & 
   {I\over {m_c }} ({ m_{\Lambda_c} \over m_{\Lambda_b}})^{1\over 2}
 exp[-\alpha \{( \beta -y)^2 - \gamma y\}] 
\nonumber\\
& { } & [m_c - m_{\Lambda_c}+
{{ 2 m_u \alpha_2^b}\over {m_{\Lambda_c}m_b(\alpha_2^c 
+\alpha_2^b)}}(m_bm_{\Lambda_c} \mp m_cm_{\Lambda_b})],
\eea
\be
a  =  
   I (4 m_{\Lambda_c} m_{\Lambda_b})^{1\over 2}
 exp[-\alpha \{( \beta -y)^2 - \gamma y\} ],
\ee
\bea
a_{\pm}  & =  & 
   {I\over m_c} {( {m_{\Lambda_c} \over m_{\Lambda_b}})}^{1\over 2} 
  exp[-\alpha \{( \beta -y)^2 - \gamma y\}]\nonumber\\
 & { } & [m_c-m_{\Lambda_c} 
+{{ 2 m_u \alpha_2^b}\over {m_b(\alpha_2^c +\alpha_2^b)}}(m_b +m_c)],
\eea
 where 
$$\alpha={{3 m_u^2 m_{\Lambda_b}^2}
\over {4 m_{\Lambda_c}^2 (\alpha_2^b +\alpha_2^c)}},$$
 $$\beta = 1 - {m_{\Lambda_c}^2\over {m_{\Lambda_b}^2}},$$
 $$\gamma = {{4 m_{\Lambda_c}^2}\over {m_{\Lambda_b}^2}},$$
having numerical values $\alpha=0.43$, $\beta=0.83$ and $\gamma=0.69$ 
for our choice of values of the parameters.

\bigskip

\centerline{\bf IV. RESULTS AND CONCLUSION}

\bigskip

In our calculation, we use nonrelativistic wave functions which gives the y 
 dependence $ exp[-\alpha \{( \beta -y)^2 - \gamma y\}]$ for all the 
form factors. The 
 form factors at zero recoil are obtained as $g = 8.02 \quad GeV$, $ g_+ =
 -0.15 $,
 $g_- = 0.075 $, 
$a= 7.07 \quad GeV$ and $a_{\pm} = 0.008$ from which inserting 
the common form factor the quantities of physical interest may be calculated. 
We get the numerical
 value of decay width and branching 
ratio (where experimental 
value of the life time of $\Lambda_b$ has been used)
as $5.7\times 10^{10} \quad sec^{-1}$ and
 $7.6\%$ respectively.

Our results compare favourably well with \cite{single,guo,korner}. 
As an example,
we list the results of Ref. \cite{single}
where in order to calculate the y dependence of the form factors, 
 use has been made of  the pole dominated form factors
  with a pole at $m^* = 6.0 \quad GeV$.
 They assume the same resonance mass for all the form factors, which yields
  the same y dependence 
for all the form factors. They have found the numerical
 values of decay width 
and branching ratio to be $5.7\times 10^{10} \quad sec^{-1}$
  and $7.7\%$ respectively.

At present, 
semileptonic decays of bottom baryons \cite{pdg} are studied mainly from
 the LEP experiments to extract lifetimes 
 from Z decays.. 
The best experimental value \cite{pdg} of $B({\Lambda}_b \to
 {\Lambda}_c l^- \nu_l X) = {{\Gamma( B({\Lambda}_b \to
 {\Lambda}_c {l^-} \nu_l X)}\over {{\Gamma}_{total}}}$ 
( most of these decays 
 are presumed to be three body decays) is $(10.0 \pm 3.0)\%$. However,
 this is not a pure measurement since what experimentalists actually 
measure is the
 product branching fraction, viz. 
$B(\Lambda_b \to \Lambda_c l^- \nu_l X) \times f(b \to \Lambda_b)$ 
where $f(b\to \Lambda_b) $ is fraction of b-flavored 
baryons produced in $Z\to b \bar b$ decays.
 So our calculated estimate  
compares reasonably well with experimental value
 taking into account the uncertainties in 
the present experimental measurements.

{\bf Acknowledgments}

One of us (BDR) is grateful to Professor Bikash Sinha and Dr. Dinesh
K. Srivastava for providing him 
facilities for research at the Variable Energy Cyclotron Centre (Calcutta).
KSG would like to thank A. Abd El-Hady, A. Datta and J. Schechter for
discussions. DC would like to acknowledge illuminating discussions 
with A. Dasgupta.

\end{document}